\def\ra{\rangle}
\def\la{\langle}
\def\be{\begin{equation}}
\def\ee{\end{equation}}
\def\ba{\begin{array}}
\def\ea{\end{array}}

\def\Cb{{\Bbb C}}

\documentclass[aps,amsmath,amssymb,amsfonts,showpacs,pra]{revtex4}
\usepackage{graphicx}
\usepackage{epstopdf}
\def\qed{\leavevmode\unskip\penalty9999 \hbox{}\nobreak\hfill
     \quad\hbox{\leavevmode  \hbox to.77778em{%
               \hfil\vrule   \vbox to.675em%
               {\hrule width.6em\vfil\hrule}\vrule\hfil}}
     \par\vskip3pt}

\begin{document}

\title{A lower bound of concurrence for multipartite quantum systems}

\author{Xue-Na Zhu$^{1}$}
\author{Ming Li$^{2}$}
\author{Shao-Ming Fei$^{3,4}$}

\affiliation{$^1$School of Mathematics and Statistics Science, Ludong University, Yantai 264025, China\\
$^2$College of the Science, China University of
Petroleum, Qingdao 266580, China\\
$^3$School of Mathematical Sciences, Capital Normal
University, Beijing 100048, China\\
$^4$Max-Planck-Institute for Mathematics in the Sciences, 04103 Leipzig, Germany}

\begin{abstract}
We present a lower bound of
concurrence for four-partite systems in terms of the concurrence for $M\, (2\leq M\leq 3)$ part quantum
systems and give an analytical lower bound for $2\otimes2\otimes2\otimes2$ mixed quantum sates.
It is shown that these lower bounds
are able to improve the existing bounds  and detect
entanglement better. Furthermore, our approach can be generalized to multipartite quantum systems.
\end{abstract}


\maketitle

\section{Introduction}\label{sec1}

 Quantifying
entanglement is a basic and long standing problem in quantum
information theory \cite{q}. Concurrence \cite{g,s6,s7,s8,Fei} is one of the well-accepted
entanglement measures \cite{t1,t2,t3,t4,t5,t6}. Different from the entanglement
of formation which is defined for bipartite systems, concurrence can be generalized
to arbitrary multipartite systems. Nevertheless, calculation of
concurrence is a formidable task for higher-dimensional cases.
For arbitrary $S$-dimensional bipartite
quantum states, Ref. \cite{zhao} provided an
analytical lower bound of concurrence by decomposing the
joint Hilbert space into many $s$ $(2\leq s\leq S-1)$-dimensional
subspaces, which may be used to improve all the known lower
bounds of concurrence.
For arbitrary  qubit systems, Ref. \cite{zhuq}
provided
analytical lower bounds of concurrence in terms of the monogamy inequality
of concurrence for qubit systems. For arbitrary N-partite $S$-dimensional quantum
states, Ref. \cite{three}
provided  an analytical
lower bound of concurrence in terms of the concurrence for N-partite $s$ $(2\leq s\leq S-1)$-dimensional quantum
systems. More generally, for arbitrary N-partite arbitrary dimensional quantum
states, Ref. \cite{liming} provided an analytical
lower bound of concurrence in terms of the concurrence for two part quantum
systems. A natural problem is whether the
arbitrary dimensional N-partite quantum states can be dealt
with $M$-partite ($2\leq M\leq N-1$) quantum systems.

In this paper we provide the lower
bound of concurrence for 4-partite quantum states in terms of tripartite and  bipartite quantum
systems. The generalized lower bound of concurrence can be
generalized to the multipartite case.

\section{Lower bound of concurrence for four-partite quantum systems}
To investigate multi-entanglement we first introduce the M-partite
concurrence in N-partite systems.
For a pure N-partite quantum state $|\psi\ra\in H_{1}\otimes H_{2}\otimes\cdots\otimes
{H}_{N}$, $dim {H}_{i}=d_i$, $i=1,...,N$,
the concurrence of $|\psi\ra$ is defined by \cite{Fei,prl}
\begin{eqnarray}\label{chun}
C_{N}(|\psi\ra)=
2^{1-\frac{N}{2}}\sqrt{(2^{N}-2)-\sum_{\alpha}Tr(\rho_{\alpha}^{2})},
\end{eqnarray}
where $\rho_{\alpha}=Tr_{\bar{\alpha}}(|\psi\ra\la\psi|)$,
$\alpha\subseteq\{1,2,...,N\}$, $\bar{\alpha}$ is the compliment of ${\alpha}$,
$\rho_{\alpha}$ labels all the different reduced density matrices of $|\psi\ra\la\psi|$. We
list all the $2^N-2$ reduced matrices in the following way:
$\{\rho_1,
\rho_2,...,\rho_N,\rho_{12},\rho_{13},...,\rho_{1N},\rho_{23},...,\rho_{12...
N-1},...,\rho_{23... N}\}$, by noticing that
$Tr(\rho_{\alpha}^{2})=Tr(\rho_{\bar{\alpha}}^{2})$ for any pure
states. For a mixed multipartite quantum state,
$\rho=\sum_{i}p_{i}|\psi_{i}\ra\la\psi_{i}| \in {\mathcal
{H}}_{1}\otimes{\mathcal {H}}_{2}\otimes\cdots\otimes{\mathcal
{H}}_{N}$, the corresponding concurrence is given by the convex roof extension,
\begin{eqnarray}\label{mixed}
C_{N}(\rho)=\min_{\{p_{i},|\psi_{i}\}\ra}\sum_{i}p_{i}C_{N}(|\psi_{i}\ra),
\end{eqnarray}
where the minimum is taken over all possible pure state
decompositions ${\{p_{i},|\psi_{i}\rangle\}}$ of $\rho$.
This multipartite concurrence can be used to detect and classify
various genuine multipartite entanglements.
It has been shown in \cite{liming} that a multipartite quantum state is genuinely multipartite entangled if the multipartite concurrence is larger
than certain quantities given by the number and the dimension of the subsystems.

For the N-partite quantum state $|\psi\ra\in {\mathcal
{H}}_{1}\otimes{\mathcal {H}}_{2}\otimes\cdots\otimes{\mathcal
{H}}_{N}$, we denote the general $M$-partite
decomposition of $|\psi\ra$,
$\{i^1\},...,\{i^{M_1}\}, \{k^1_1,k^1_2\},
...,\{k^{M_2}_1,k^{M_2}_2\},...,\{q^{M_j}_1,...,q^{M_j}_j\}$,
where all the subspaces in one bracket $\{\}$ are taken to be
one part. Hence, $\{i^1,...,i^{M_1}, k^1_1,k^1_2,...,k^{M_2}_1,k^{M_2}_2,...,q^{M_j}_1,...,
q^{M_j}_j\}=\{1,2,...,N\}$ and  $\sum_{k=1}^{j}M_k=M$, $\sum_{k=1}^{j}kM_k=N$.
The concurrence of the state $|\psi\ra$ under such $M$-partite partition is given by
\begin{eqnarray}\label{CM}
C_{M}(|\psi\ra)=2^{1-\frac{M}{2}}\sqrt{(2^{M}-2)
-\sum_{\beta}Tr(\rho_{\beta}^{2})},
\end{eqnarray}
where $\beta\in\{ \{i^1\},...,\{i^{M_1}\},...,\{q^{M_j}_1,...,q^{M_j}_j\} \}.$
Take $N=4$ and $M=3$ as an example, one has $M_1=2$ and $M_2=1$.
There are six different tripartite decompositions:
$1|2|34$, $1|23|4$, $1|24|3$, $12|3|4$, $13|2|4$ and $14|2|3$.
For convenience, we denote $C_{i|jk|l}(\rho)=C_3(\rho_{i|jk|l})$
and $C_{ij|kl}(\rho)=C_2(\rho_{ij|kl})$.

{\bf{Theorem 1:}} For any mixed quantum state, $\rho
\in {\mathcal {H}}_{1}\otimes{\mathcal
{H}}_{2}\otimes{\mathcal {H}}_{3}\otimes{\mathcal
{H}}_{4}$, the concurrence is bounded by
\begin{eqnarray}\label{xia}
C^2_{4}(\rho) &\geq& \frac{1}{12}\big(2C^2_{1|2|34}(\rho)
+2C^2_{1|3|24}(\rho) \nonumber\\
&+&2C^2_{1|4|23}(\rho)+2C^2_{12|3|4}(\rho)\nonumber\\
&+&
2C^2_{13|2|4}(\rho) +2C^2_{14|2|3}(\rho)\nonumber\\
&+&C^2_{12|34}(\rho) +C^2_{13|24}+C^2_{14|23}(\rho) \big).
\end{eqnarray}

[Proof:] We start the proof with a pure state $|\psi\ra\in {\mathcal
{H}}_{1}\otimes{\mathcal {H}}_{2}\otimes{\mathcal
{H}}_{3}\otimes{\mathcal
{H}}_{4}$. According to the definition (\ref{chun}), one has that
\begin{eqnarray}\label{proof1}
C_{4}(|\psi\ra)&=&\frac{1}{2}\sqrt{14-\sum_{\alpha}
Tr(\rho_{\alpha}^{2})}\nonumber\\
&=&\frac{1}{2}\sqrt{\sum_{\alpha}\left(1-Tr(\rho_{\alpha}^{2})\right)},
\end{eqnarray}
where $\rho_{\alpha}$ labels all the different reduced density matrices of $|\psi\rangle\langle\psi|$.
On the other hand, we have
\begin{eqnarray}\label{11}
C^2_{i|jk|l}(|\psi\rangle)
&=&\frac{1}{2}\big[(1-Tr(\rho^2_{i}))
+(1-Tr(\rho^2_{jk}))\nonumber\\
&+&(1-Tr(\rho^2_{l}))
+(1-Tr(\rho^2_{ijk}))\nonumber\\
&+&(1-Tr(\rho^2_{il}))+(1-Tr(\rho^2_{jkl}))\big],
\end{eqnarray}
and
\begin{eqnarray}\label{12}
C^2_{ij|kl}(|\psi\rangle)=(1-Tr(\rho^2_{ij}))+(1-Tr(\rho^2_{kl})).
\end{eqnarray}

From (\ref{proof1}), (\ref{11}) and (\ref{12}) , we have
 \begin{eqnarray}
C^2_{4}(|\psi\ra)&=& \frac{1}{12}\big
(2C^2_{1|2|34}(|\psi\rangle)+2C^2_{1|3|24}(|\psi\rangle)\nonumber\\
&+&2C^2_{1|4|23}(|\psi\rangle)
+2C^2_{12|3|4}(|\psi\rangle)\nonumber\\
&+&2C^2_{13|2|4}(|\psi\rangle)
+2C^2_{14|2|3}(|\psi\rangle)\nonumber\\
&+&C^2_{12|34}(|\psi\rangle)+C^2_{13|24}(|\psi\rangle)+C^2_{14|23}(|\psi\rangle)\big).
\end{eqnarray}
Let $\rho=\sum_ip_i|\psi\rangle_i\langle\psi|$ be the
optimal pure state decomposition of (\ref{mixed}). We have
\begin{widetext}
\begin{eqnarray}
C_4(\rho)
&=&\sum_{i}p_iC_{4}(|\psi\rangle_i)\nonumber\\
&=&\sum_{i}p_i\sqrt{\frac{1}{6}C^2_{1|2|34}(|\psi\rangle_i)+...
+\frac{1}{6}C^2_{14|2|3}(|\psi\rangle_i)
+\frac{1}{12}C^2_{12|34}(|\psi\rangle_i)+...
+\frac{1}{12}C^2_{14|24}(|\psi\rangle_i)}\nonumber\\
&\geq&\sqrt{(\sum_ip_i\frac{1}{\sqrt{6}}C_{1|2|34}(|\psi\rangle_i))^2+
...+(\sum_ip_i\frac{1}{\sqrt{12}}C_{14|23}(|\psi\rangle_i))^2}\nonumber\\[1mm]
&\geq&\frac{1}{\sqrt{12}}\big(2C^2_{1|2|34}(\rho)+2C^2_{1|3|24}(\rho)
+2C^2_{1|4|23}(\rho)+2C^2_{12|3|4}(\rho)
+2C^2_{13|2|4}(\rho)+2C^2_{14|2|3}(\rho)\nonumber\\[2mm]
&&+C^2_{12|34}(\rho)+C^2_{13|24}(\rho)+C^2_{14|23}(\rho)\big)^{\frac{1}{2}},
\end{eqnarray}
\end{widetext}
where the Cauchy-Schwarz inequality $(\sum_j (\sum_i y_{ij})^2 )^{\frac{1}{2}} \leq
\sum_i (\sum_j y_{ij}^2)^{\frac{1}{2}}$ has been used in the second
inequality.
\qed

For a mixed quantum state $\rho\in H_1\otimes H_2\otimes H_3\otimes H_4,$
a lower bound of $C^2_{4}(\rho)$ has been derived based on bipartite
partitions in Ref. \cite{liming}. By using the following relation \cite{liming},
\begin{eqnarray}
C^2_{i|j|kl}(\rho)\geq \frac{1}{2}\left(C^2_{i|jkl}(\rho)+C^2_{ij|kl}(\rho)+C^2_{ikl|j}(\rho)\right),
\end{eqnarray}
from (\ref{xia}) we have
\begin{widetext}
\begin{eqnarray}
C^2_{4}(\rho) &\geq& \frac{1}{12}\big(2C^2_{1|2|34}(\rho) +2C^2_{1|3|24}(\rho)
+2C^2_{1|4|23}(\rho) +2C^2_{12|3|4}(\rho)
+2C^2_{13|2|4}(\rho) +2C^2_{14|2|3}(\rho) \nonumber\\
&+&C^2_{12|34}(\rho) +C^2_{13|24}+C^2_{14|23}(\rho) \big)\nonumber\\
&\geq&\frac{1}{4}\big(C^2_{1|234}(\rho) +C^2_{2|134}(\rho) +C^2_{3|124}(\rho)
+C^2_{4|123}(\rho)\nonumber\\
&+&C^2_{12|34}(\rho) +C^2_{13|24}(\rho)+C^2_{14|23}(\rho)\big)=\Delta,
\end{eqnarray}
\end{widetext}
where $\Delta$ is the lower bound obtained in \cite{liming}. Hence, our bound (\ref{xia})
is better than the lower bound in \cite{liming} for four-partite quantum mixed states.

\section{Analytical lower bound for $2\otimes2\otimes2\otimes2$ mixed states}
Let $H_{A} $, $H_{B}$ and $H_{C} $ be  $2,2$ and $4$-dimensional
Hilbert spaces associated with the systems $A$, $B$ and $C$, respectively. A
pure state $\vert\varphi\rangle\in H_{A}\otimes H_{B}\otimes
H_{C}$ has the form
\begin{equation}\label{ij}
\vert\varphi\rangle=\sum_{i=0}^{1}\sum_{j=0}^{1}\sum_{k=0}^{3}a_{ijk}\vert
ijk\rangle,
\end{equation}
where $a_{ijk}\in\Cb$,
$\sum_{ijk}|a_{ijk}|^2=1$, $\{\vert ijk\rangle\}$ is
the basis of $H_{A}\otimes H_{B}\otimes H_{C}$.
The concurrence of $\vert\varphi\rangle$ can be
equivalently written as \cite{Fei},
\begin{widetext}
\begin{eqnarray}\label{deng}
C_{3}(|\varphi\rangle)=\sqrt{\frac{1}{2}\sum(|a_{ijk}a_{pqm}-a_{ijm}a_{pqk}|^2
+|a_{ijk}a_{pqm}-a_{iqk}a_{pjm}|^2+|a_{ijk}a_{pqm}-a_{pjk}a_{iqm}|^2)}.
\end{eqnarray}
\end{widetext}

To evaluate $C_3(\rho)$, we project $2\otimes2\otimes4$ dimensional states
 to $2\otimes2\otimes2$ sub-states.
For a given $2\otimes 2\otimes 4$ pure
state, we define its ``$2\otimes 2\otimes 2$" pure state
$|\varphi\rangle_{2\otimes 2\otimes 2}=
\sum_{i=0}^{1}\sum_{j=0}^{1}\sum_{k\in\{k_1,k_2\}}a_{ijk}|ijk\rangle, $
where $\{k_1,k_2\}\in\{\{0,1\},\{0,2\},\{0,3\},\{1,2\},\{1,3\},\{2,3\}\}$.
In fact, for any $2\otimes2\otimes4$ pure state $|\varphi\rangle$ , there are $6$
different $2\otimes2\otimes2$ substates with respect to $|\varphi\rangle$.
Without causing confusion, in the following we simply use $|\varphi\rangle_{2\otimes 2\otimes 2}$
to denote one of such states, as these substates will always be
considered together.
The concurrence $C(|\varphi\rangle_{2\otimes 2\otimes 2})$ is similarly
given by Eq. (\ref{deng}), with the subindices $i$ and $j$, associated with the systems $A$ and $B$ respectively,
running from $0$ to $1$, and with the subindex $k$ associated with the system $C$ taking values $k_1$ and $k_2$

Correspondingly, for a mixed state $\rho$, we define its
$2\otimes 2\otimes 2$ mixed (unnormalized) substates $\rho_{2\otimes 2\otimes 2}$.
The concurrence of
$\rho_{2\otimes 2\otimes 2}$ is defined by
$C(\rho_{2\otimes 2\otimes 2})=\min\sum_ip_iC(|\phi_i\rangle),$
minimized
over all possible $2\otimes2\otimes2$ pure-state decompositions of
$\rho_{2\otimes 2\otimes 2}=\sum_ip_i|\phi_i\rangle\langle\phi_i|,$
with$\sum_ip_i=Tr(\rho_{2\otimes 2\otimes 2})$.
The $2\otimes 2 \otimes 2$ submatrices $\rho_{2\otimes 2\otimes 2}$ have the following form,
\begin{widetext}
\begin{eqnarray}
 \rho_{2\otimes 2\otimes 2}=\begin{pmatrix}
   \rho_{00k_1,00k_1} & \rho_{00k_1,00k_2} & \rho_{00k_1,01k_1}&
   \rho_{00k_1,01k_2}& \rho_{00k_1,10k_1}
   & \rho_{00k_1,10k_2}  & \rho_{00k_1,11k_1}& \rho_{00k_1,11k_2}\\
     \rho_{00k_2,00k_1} & \rho_{00k_2,00k_2} & \rho_{00k_2,01k_1}
     & \rho_{00k_2,01k_2}& \rho_{00k_2,10k_1}
   & \rho_{00k_2,10k_2}  & \rho_{00k_2,11k_1}& \rho_{00k_2,11k_2}\\
  \rho_{01k_1,01k_1} & \rho_{01k_1,00k_2} & \rho_{01k_1,01k_1}&
   \rho_{01k_1,01k_2}& \rho_{01k_1,10k_1}
   & \rho_{01k_1,10k_2}  & \rho_{01k_1,11k_1}& \rho_{01k_1,11k_2}\\
     \rho_{01k_2,00k_1} & \rho_{01k_2,00k_2} & \rho_{01k_2,01k_1}
     & \rho_{01k_2,01k_2}& \rho_{01k_2,10k_1}
   & \rho_{01k_2,10k_2}  & \rho_{01k_2,11k_1}& \rho_{01k_2,11k_2}\\
   \rho_{10k_1,00k_1} & \rho_{10k_1,00k_2} & \rho_{10k_1,01k_1}&
   \rho_{10k_1,01k_2}& \rho_{10k_1,10k_1}
   & \rho_{10k_1,10k_2}  & \rho_{10k_1,11k_1}& \rho_{10k_1,11k_2}\\
     \rho_{10k_2,00k_1} & \rho_{10k_2,00k_2} & \rho_{10k_2,01k_1}
     & \rho_{10k_2,01k_2}& \rho_{10k_2,10k_1}
   & \rho_{10k_2,10k_2}  & \rho_{10k_2,11k_1}& \rho_{10k_2,11k_2}\\
   \rho_{11k_1,00k_1} & \rho_{11k_1,00k_2} & \rho_{11k_1,01k_1}&
   \rho_{11k_1,01k_2}& \rho_{11k_1,10k_1}
   & \rho_{11k_1,10k_2}  & \rho_{11k_1,11k_1}& \rho_{11k_1,11k_2}\\
     \rho_{11k_2,00k_1} & \rho_{11k_2,00k_2} & \rho_{11k_2,01k_1}
     & \rho_{11k_2,01k_2}& \rho_{11k_2,10k_1}
   & \rho_{11k_2,10k_2}  & \rho_{11k_2,11k_1}& \rho_{11k_2,11k_2}\\
  \end{pmatrix},\quad
\end{eqnarray}
\end{widetext}
where $0\leq k_1<k_2\leq3$ associated to the space $H_{C}$.

{\bf{Theorem 2:}}\label{th2}
For any $2\otimes 2 \otimes 4$  tripartite mixed quantum
state $\rho$, the concurrence $C(\rho)$ satisfies
\be\label{1.6}
C_3^2(\rho) \geq \frac{1}{3} \sum C_3^2(\rho_{2\otimes 2\otimes 2}),
\ee
where
$\sum$ stands for summing over all possible $2\otimes 2\otimes 2$ mixed sub-states $\rho_{2\otimes 2\otimes 2}$.

[Proof]. From the expression of concurrence (\ref{deng}), it is straightforward to prove that the concurrence of
pure state $\vert\varphi\rangle$ and  the concurrence of
$|\varphi\rangle_{2\otimes 2\otimes 2}$ with respect to $\vert\varphi\rangle$ have the following relation,
\begin{eqnarray}
C_3^2(\vert\varphi\rangle)
\geq\sum\frac{1}{3}C_3^2(|\varphi\rangle_{2\otimes 2\otimes 2}).
\end{eqnarray}
Therefore for mixed state $\rho=\sum p_i |\varphi_i\rangle \langle \varphi_i|$, we have
\begin{eqnarray*}
\ba{rcl}
C_3(\rho)
&=&\displaystyle\min \sum_i p_i C_3(|\varphi_i\rangle)\\[1mm]
&\geq& \displaystyle\min \sqrt{\frac{1}{3}}\sum_i p_i \left(\sum
C_3^2(|\varphi_i\rangle_{2\otimes 2\otimes 2})\right)^{\frac{1}{2}}\\[1mm]
&\geq&\displaystyle \min \sqrt{\frac{1}{3}}\left[\sum \left(\sum_i p_i C_3(|\varphi_i\rangle_{2\otimes 2\otimes 2})
\right)^2\right]^{\frac{1}{2}}\\[1mm]
&\geq&\displaystyle\sqrt{\frac{1}{3}} \left[\sum \left(\min \sum_i p_i C_3(|\varphi_i\rangle_{2\otimes 2\otimes 2})
\right)^2\right]^{\frac{1}{2}}\\[1mm]
&=&\displaystyle\sqrt{\frac{1}{3}} \left[\sum C_3^2(\rho_{2\otimes 2\otimes 2})\right]^{\frac{1}{2}},
\ea
\end{eqnarray*}
where the relation $(\sum_j (\sum_i x_{ij})^2 )^{\frac{1}{2}} \leq
\sum_i (\sum_j x_{ij}^2)^{\frac{1}{2}}$ has been used in the second
inequality, the first three minimizations run over all possible pure
state decompositions of the mixed state $\rho$, while the last
minimization runs over all $2\otimes 2 \otimes 2$ pure state decompositions
of $\rho_{2\otimes 2\otimes 2}$ associated with $\rho$. \qed

According to Theorem 1 and Theorem 2,
we have the following Corollary 1:

{\bf{Corollary 1:}}
For any $2\otimes 2 \otimes 2\otimes 2$   mixed quantum
state $\rho$, the concurrence $C_4(\rho)$ satisfies
\be\label{1.6}
C_4^2(\rho) \geq  \frac{1}{12}\left
(\sum\frac{2}{3}C_3^2(\rho_{2\otimes2\otimes2})+\sum_{ 2\leq j\leq4}C^2_{1j|\{1,2,3,4\}\setminus\{1,j\}}(\rho)\right),
\ee
where
$\sum$ stands for summing over all possible $2\otimes 2\otimes 2$ mixed sub-states $\rho_{2\otimes 2\otimes 2}$ of $\rho_{i|j|kl}$, $1\leq i<j\leq4$, $\{k,l\}=\{1,2,3,4\}\setminus\{i,j\})$.

For any four-qubit mixed quantum state $\rho$, Ref. \cite{zhuq} provided analytical
lower bounds of concurrence in terms of the monogamy inequality of concurrence:
\begin{equation}
C_4^2(\rho) \geq \frac{1}{2}\sum_{i=1}^{3}\sum_{j>i}^{4}(T_i+T_j)C^2_{ij}(\rho),
\end{equation}
 where $T_i$ $(i=1,2,3,4)$ are given in Ref. \cite{zhuq} and the difference of a constant factor
 $\frac{1}{2}$ defining the concurrence for four qubit pure states has already been taken into account.
The bounds given in Corollary 1 can be used to
improve the bounds of concurrence presented in \cite{zhuq}. Let us consider the following  example.

{\it Example:} We consider the quantum state
$\rho=\frac{1-t}{16}I_{16}+t|\psi\rangle\langle\psi|$,
where $|\psi\rangle=({|0000\rangle+|0011\rangle+|1100\rangle+|1111\rangle})/{2}$,
$I_{16}$ is the $16\times 16$ identity matrix.

From our Theorem 1, we need to compute the lower bounds of $C_{i|j|kl}(\rho)$
and  $C_{ij|kl}(\rho)$.
For convenience, we denote
\begin{equation}\nonumber
Z_1=\frac{(5t-1)^2}{128},
\end{equation}
\begin{equation}\nonumber
Z_2=\frac{1}{128}(1-9t)^2(1+t)^2,
\end{equation}
\begin{equation}\nonumber
Z_3=-\frac{1}{256}(1+t)^2\left[5(-51+4\sqrt{17})t^2+(26+4\sqrt{17})t-3\right],
\end{equation}
and
\begin{equation}\nonumber
Z_4=\frac{3(1+t)^4}{128}\left(\sqrt{\frac{1+7t}{4(t+1)}}
-3\sqrt{\frac{1-t}{4(t+1)}}\right)^2.
\end{equation}
For $C(\rho_{i|j|kl})$, we use Theorem 2 and the lower bound of \cite{three} of
 $2\otimes2\otimes2$ mixed states.
We obtain the lower bound $Z$ of
$\sum_{1\leq i<j\leq4,\{k,l\}=\{1,2,3,4\}/\{i,j\}}C_{i|j|kl}^2(\rho)$:
\begin{equation}\nonumber
Z=\left\{
\begin{aligned}
&2Z_2,~~~~~~~~~~~~~~~~~~ t\in(\frac{1}{9},0.2],\\
&32 Z_1+2Z_2,~~~~~~~~t\in(0.2,0.308051],\\
&32 Z_1+Z_2+Z_3,~~~t\in(0.308051,1].
\end{aligned}
\right.
\end{equation}

For $C(\rho_{ij|kl})$, we use the lower bound of \cite{zhao}.
We have
$$\sum_{1<j\leq4,\{k,l\}=\{1,2,3,4\}\setminus\{1,j\}}C_{1j|kl}^2(\rho)
\geq Z_4$$ with $t\in(0.5,1]$.

Then the lower bound of $C_4^2(\rho)$
can be obtained, see Fig. 1.
\begin{figure}[htpb]
\renewcommand{\figurename}{Fig.}
\centering
\includegraphics[width=7.5cm]{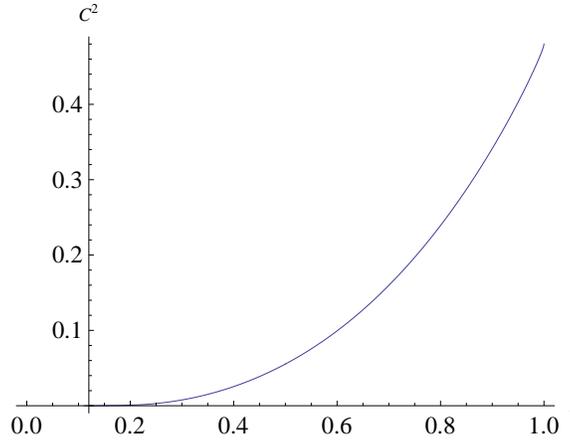}
\caption{{\small Lower bound of $C_4^2(\rho)$ for $0\leq t\leq1$ .}}
\label{1}
\end{figure}

From Fig. 1,
we see that the lower bound can detect entanglement of $\rho$ for $t>\frac{1}{9}$.
From Fig. 2, we see that the our result is better than
the lower bound from \cite{liming} and \cite{zhuq}
for $t\in(\frac{1}{9},0.4)$, where the difference of a constant factor $\frac{1}{2}$  in defining
the concurrence for four qubit pure  states has already been taken into account.
\begin{figure}[htpb]
\renewcommand{\figurename}{Fig.}
\centering
\includegraphics[width=7.5cm]{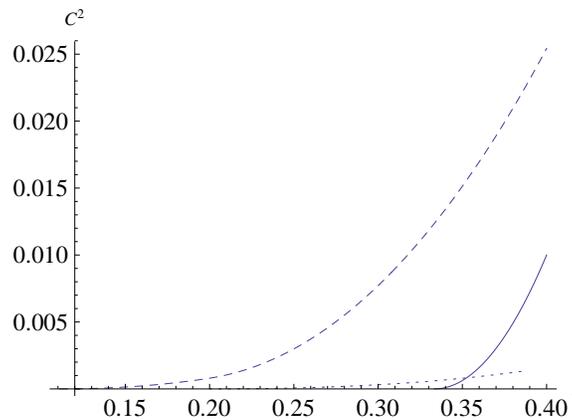}
\caption{{\small Dashed line for the bound (\ref{xia}), dotted line for that from Ref. \cite{liming},
solid line for that from Ref. \cite{zhuq}.}}
\label{2}
\end{figure}

By generalizing our results to arbitrary dimensional $n$-partite systems, we have

{\bf{Corollary 2:}}\label{c}
For any $N$-partite arbitrary dimensional mixed state
$\rho\in H_{1}\otimes H_{2}\otimes \cdots\otimes H_{N}$,
\begin{equation}
C_N^2(\rho)\geq\sum_{m=2}^{M} q_m\sum_{i}C^2_{i^1_1i^1_2...i^1_{k_1}|i^2_1i^2_2...i^2_{k_2}|
...|i^m_1i^m_2...i^m_{k_m}}(\rho),
\end{equation}
where $2\leq M\leq N-1$, $\sum_{i=1}^{m}k_m=N$, $\{i^1_1,...,i^1_{k_1},...,
i^m_1i^m_2...i^m_{k_m}\}=\{1,2,...,n\}$ and  $q_m$ is a fixed number
depending on $k_1,k_2,\cdots,k_m$.

Corollary 2 says that the lower bound of the concurrence of an N-partite quantum state can be expressed by the concurrences of its $2,3,...,N-1$-partite substates.

\section{Conclusions and Remarks}\label{sec5}
In summary, we have proposed a new approach in constructing hierarchy of lower
bounds of concurrence for mixed multipartite quantum states in terms of the less part decomposed
quantum systems.
Besides, our approach can be generalized to $N$ part systems
to obtain the lower bound of the concurrence for $M\, (2\leq M\leq N-1)$  part systems.

\bigskip
\noindent{\bf Acknowledgments}\, \,
This work is supported by NSFC under numbers 11675113 and 11605083.

\end{document}